# Novel Boron Based Multilayer Thermal Neutron Detector

M. SCHIEBER, O. KHAKHAN*
*Department of Applied Physics, the Hebrew University of Jerusalem, Jerusalem 91904 Israel*

The detector contains four or more layers of natural Boron absorbing thermal neutrons. Thickness of a layer is 0.4 - 1.2 $mg/cm^2$. The layers are deposited on one or on both sides of a metal surface used as contacts. Between the absorbing layers there are gas-filled gaps 3 - 6 mm thick. Electric field of 100 – 200 V/cm is applied to the gas-filled gaps. Natural Boron contains almost 20% of $^{10}B$ isotope. When atoms of $^{10}B$ capture a thermal neutron, nuclear reaction occurs, as a result of which two heavy particles – alpha particle and ion $^7Li$ – from the thin absorber layer are emitted in opposing sides. One of the two particles penetrates into gas-filled gap between Boron layers and ionizes the gas. An impulse of electric current is created in the gas-filled gap actuated by the applied electric field. The impulse is registered by an electronic circuit. We have made and tested detectors containing from two to sixteen layers of natural Boron with an efficiency of thermal neutron registration from 2.9% to 12.5% accordingly.

## 1. Introduction

Multilayer thermal neutron detector can be used for the real time digital image acquisition in a situation where thermal neutron stream penetrates through an object. Also the detector can be used as either thermal neutron counter, or fast neutron counter in a case where the detector is installed inside a moderator, e.g. in polyethylene. The detector has advantages over others thermal neutron detectors we are aware of: semiconductor detectors, scintillation detectors, detectors using GEM (gas electron multiplier). Our detector is simpler, inexpensive, and can have unlimited work area. The efficiency of multilayer detector increases with the number of layers converting thermal neutrons. The thermal neutron registration efficiency of a detector made of thin layers of $^{10}B$ isotope used as convertor can reach 50%, which is higher than the efficiency of tube detectors coated from inside with a thin layer of solid $^{10}B$ with a wire as a central electrode. The efficiency of those detectors is not higher than 14%. In addition, as opposed to many other gas containing detectors, our detector is flat, which makes it much more convenient to use.

Because of Helium-3 gas reserve depletion in the world, Boron based multilayer thermal neutron detector can successfully replace Helium-3 gas containing neutron detectors. Our detector can be used also for image acquisition of objects containing inflammables and explosives (in solid and liquid forms), even if the object is located inside a metal container. Our detector can be used for neutron radiation detection in industry or at public transportation check points like ports etc.

Our detector sensitivity to gamma-rays is of the same order of magnitude as Helium-3 gas containing detectors. This allows for high signal noise (S/N) ratio, because gamma-rays usually interfere with neutron registration.

Our detectors can have large work area, which can reach several square meters. For this assessment we proposed a brand new construction of the detector, in which every third (central) metallic electrode is composed from separated isolated pixels with an area from 0.1 to 200 $cm^2$. This allowed limiting electric capacitance of every pixel by a few pF, which is the same value as for cylindrical form detectors. Electric signal from each separate pixel is fed into its own input on the readout circuit.

## 2. Detection mechanism

Multilayer thermal neutron detector is an assembly of thin parallel metal plates mounted one above another in an isolating frame. Gaps between plates are about 3 – 6 mm. The gaps are filled with gas (e.g. air, nitrogen, argon, butane). Plate surfaces of gas-filled gaps are covered with thin layers of the neutron absorber containing $^{10}B$ isotope. The thickness of the gas layer is chosen in accordance with an average range of alpha particle having energy of up to 1.8 MeV in the gas. It is assumed that the exit angel of alpha particles to the surface of the converter is about 45 degrees. Thickness of the absorber layers is chosen in a range of 0.4 – 1.2 $mg/cm^2$. Thickness of the Boron layer should not be greater than the range of alpha particles with the energy of 1.47 MeV. Otherwise, some part of alpha particles will not be able to reach for gas layer and ionize the gas. Electric field is applied to metallic plates. The field density is about 100 – 200 V/mm. Figure 1 shows the schematic circuit of the described detector.

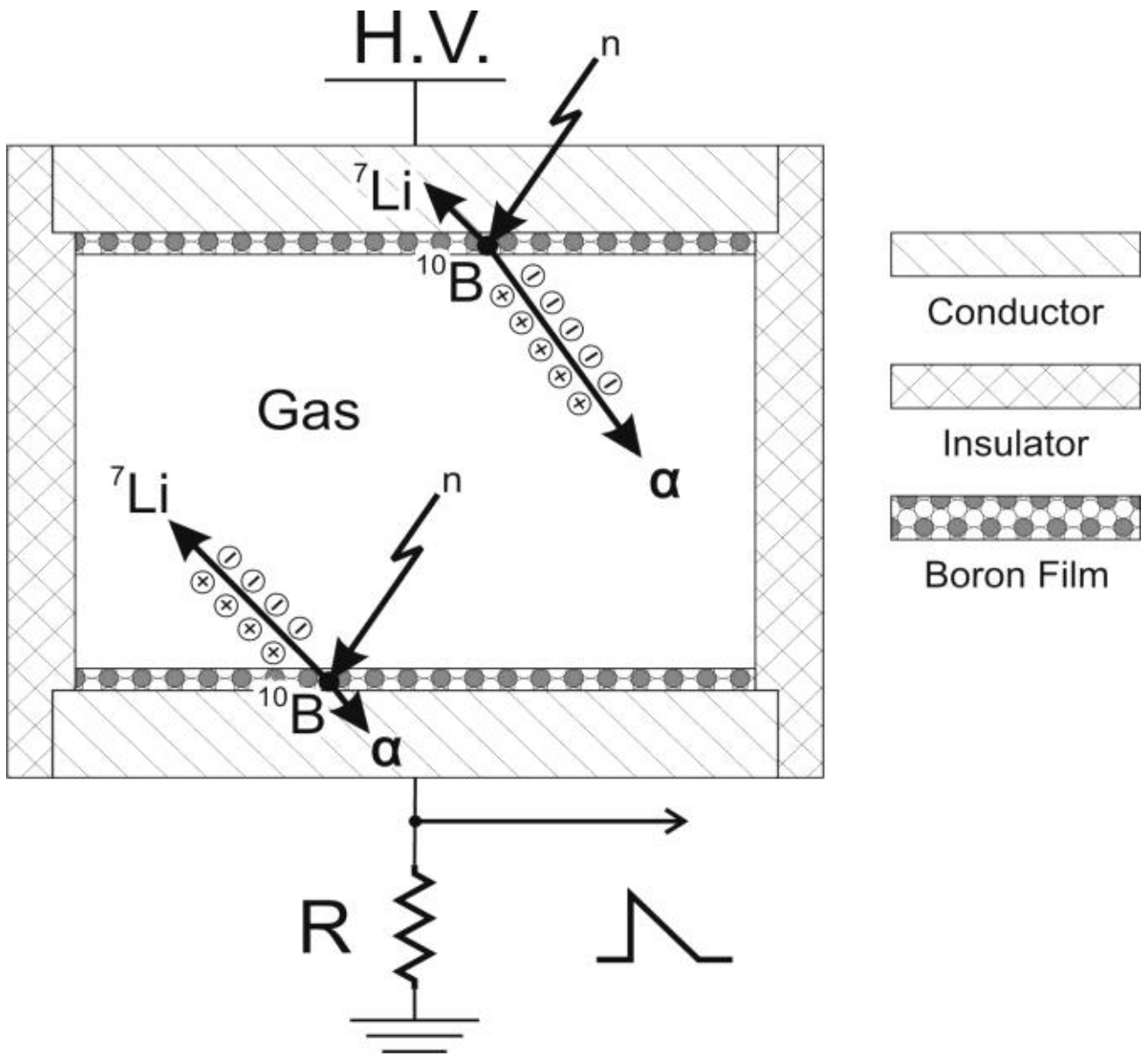

*Figure 1. Principal schema of creating signal in 2-layers boron-based thermal neutron detector***.**

The detector works as follows:
Thermal neutrons are captured by atoms of $^{10}B$, which are present in the thin layers of neutron absorbers, because $^{10}B$ has large cross-section, which is equal to 3840 barn for thermal neutrons (0.025eV);
Atom of $^{10}B$ captures a neuron, resulting in the following nuclear reaction [1]:

$$^{10}B + n \rightarrow {}^{7}Li + \alpha + \gamma$$

The following particles are created as an outcome of this reaction: alpha particles with the energy of 1.47 MeV in 94% of cases and with the energy of 1.78 MeV in 6% of cases, $^7Li$ ions with the energy of 0.84 Mev (94%) and 1.01 MeV (6%), and gamma quantum with the energy of 0.48 MeV (94%). Alpha particles and $^{7}$Li ions come apart in opposite directions according to the momentum conservation law. Therefore, an alpha particle or a $^{7}$Li ion is emitted in the direction of the gas-filled gap upon each neutron capture in a thin layer of the neutron absorber. The emitted alpha or ion ionizes corresponding layer of the gas. That is, the thin layers of neuron absorbers act as converters of neutrons into electrically charged particles. The electric field generates an electric discharge after ionization in the gas-filled gap. The resultant electric pulses are counted by the electronic readout circuit.

Gamma-quantum produced in neutron capture by an atom of $^{10}$B hardly ionizes the gas-filled gap. Extremely low sensitivity to gamma radiation, which is always accompanied by the neutron emission, is an advantage of our detector. Low sensitivity of our detectors to the gamma radiation was tested in work [2].

## 3. Experimental arrangement

We have manufactured and tested multilayer detectors, where layers of natural Boron containing almost 20% of $^{10}$B isotope were used as the neutron absorbers. For a basic element of design we have adopted 4-layers detector which is represented on Figure 2. The increase in the number of layers of the detector is achieved by combining the basic elements in the stack (sandwich). Neutron detection efficiency increases with the number of convertor layers in the detector.

The basic 4-layers element consists of three parallel plates mounted one above the other in the isolating frame with a space between the plates of about 3 – 6 mm. The top and bottom plates can be made of a metal or of a plastic coated on the inside by a metal foil. The metal foil on the inside of the plates is coated with thin layers of thermal neutron absorber consisting of natural Boron. Plate center is made of plastic. The plate is covered from top and from bottom by a copper foil. These layers of foil on top and at bottom of the central plate are divided into four isolated regions (pixels). The area of each of these pixels is 23 $cm^2$.

Over contacting parts of the foil, the central plate is also covered from top and from bottom by two thin layers of natural Boron. Layers of natural Boron have a thickness of 1.1 $mg/cm^2$. For thin layers of natural Boron, we use a fine powder with a particle size of about 1 micron. The powder is deposited on the metal surface from a liquid mechanical mixture of powder and toluene. After drying, we obtain the thin layers of natural Boron with a given thickness.

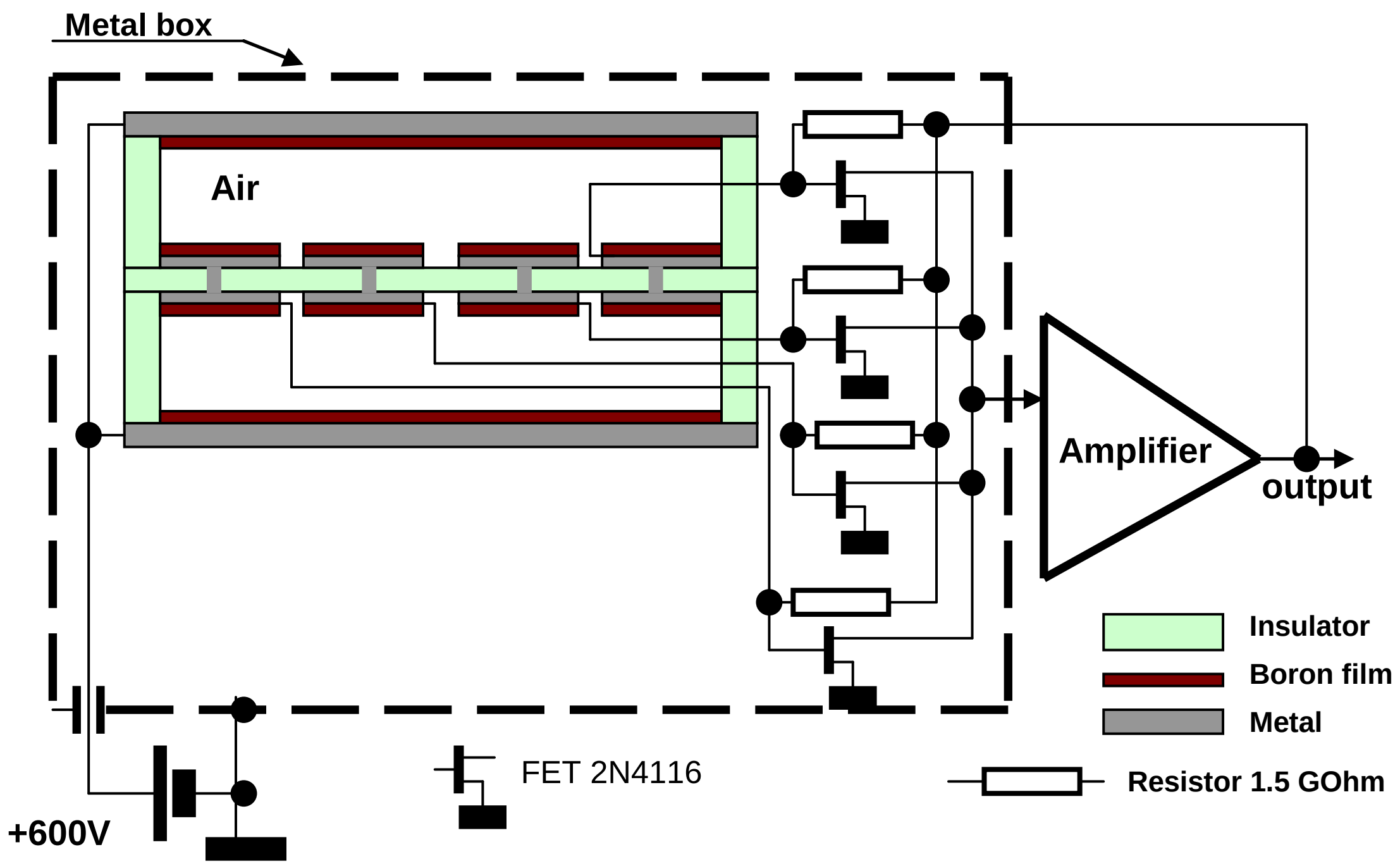

*Figure 2. Basic element 4-layers thermal neutron detector and connection schema.*

Submicron particles of powder are firmly held on a flat metal surface due to good adhesion.
High voltage is connected to the upper and to the lower metal plates. This voltage creates an electric field of the order of 100 – 200 V/mm in air gaps between the outer and central plates.

Each pair of the metal foil sections (pixels) located above and below the central plate is connected to a separate input of the preamplifier built on low-noise field-effect transistors (2N4416). The signals from the preamplifier are fed into the spectroscopy amplifier with the shaping time 10 μs, and then enter the MCA, which contains the threshold discriminator, filtering out electronic circuit noises. To protect against electromagnetic interference the detector together with the preamplifier are placed in a metal box.
The neutron source is installed on the metal box (see Figure 3).

In our work we use a radioactive source $^{241}$Am + Be, $^{241}$Am activity is 100 mCi. This source emits about 220,000 fast neutrons per second in 4P steradian. The $^{241}$Am + Be source has a ring form with a diameter of 4 cm. It is inside a lead container with a wall thickness more than 2 mm. The walls of the lead container are almost completely absorb the gamma quanta of energies up to 60 KeV emitted by the $^{241}$Am isotope.
The metal box with a detector inside is installed on a paraffin block with dimensions 10 x 24 x 24 cm (see Photo 1). The neutron source is installed on top of the metal box (see Figure 3). The container with the $^{241}$Am + Be source is placed inside the second upper paraffin block with dimensions 10 x 24 x 24 cm. The upper paraffin block is located over the metal box with the detector.

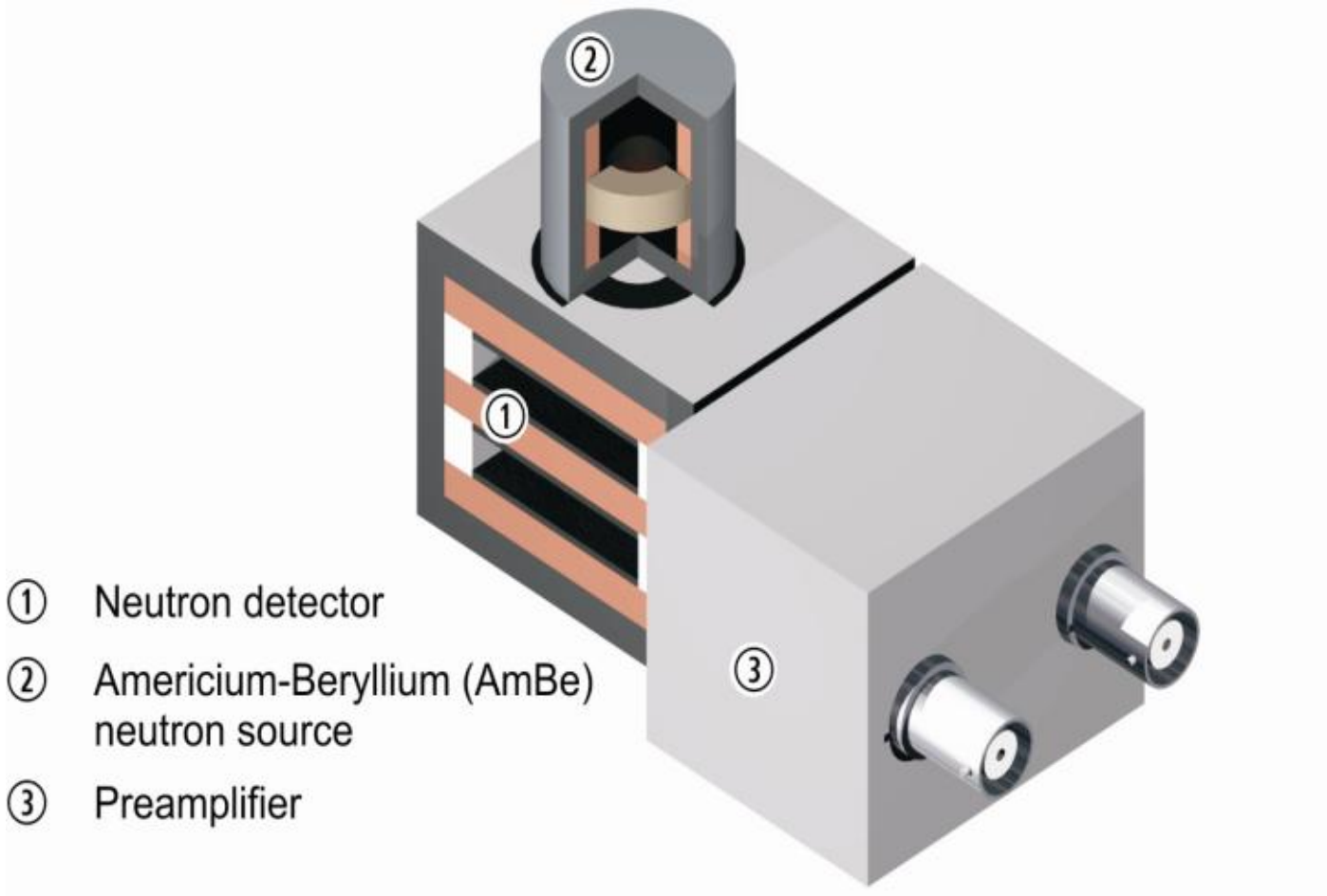

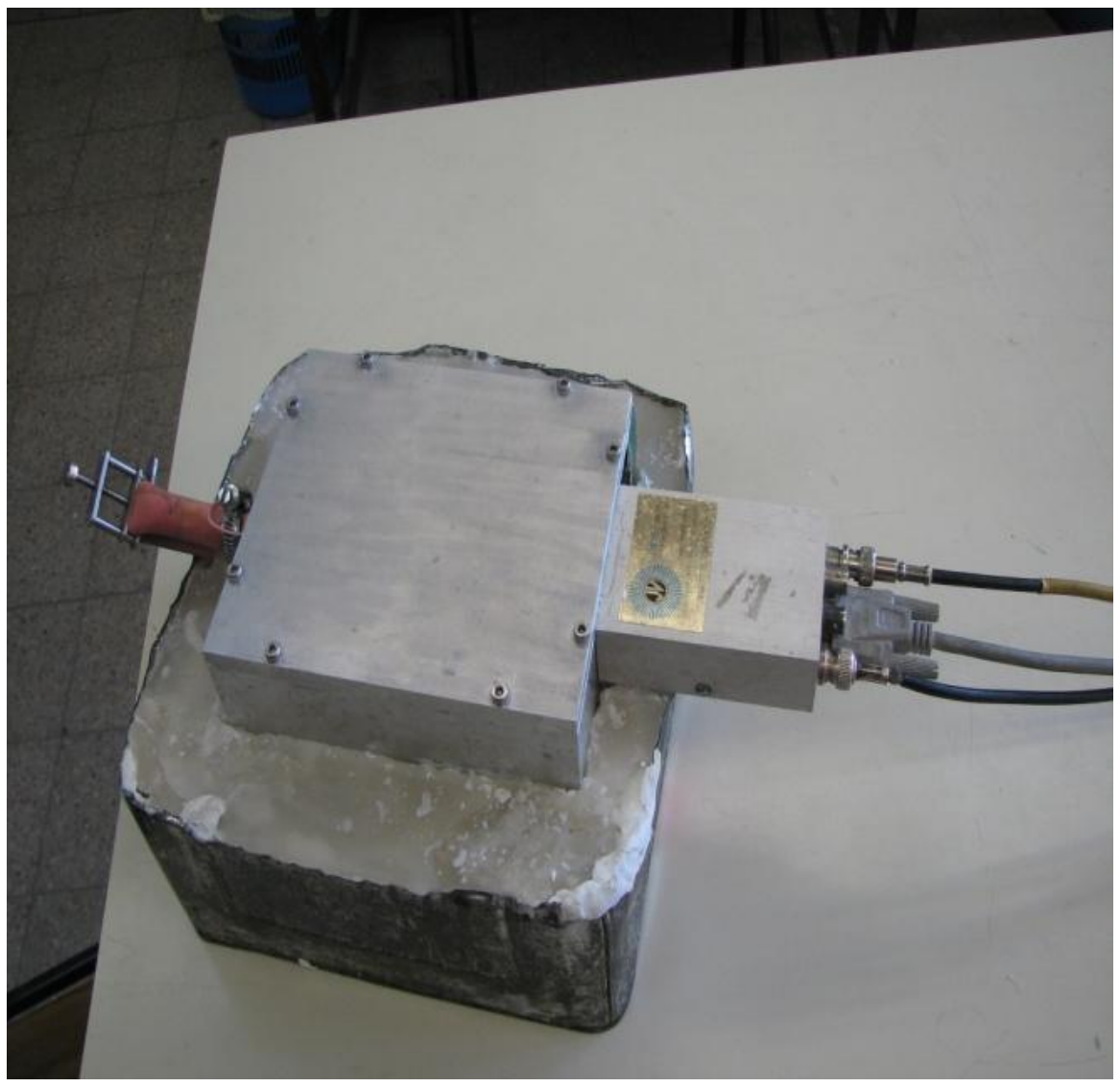

*Figure 3 Measurement setup.* *Photo1. Hermetic metal box and amplifier on bottom paraffin block.*

Fast neutrons slow-down occurs in the paraffin blocks. As a result, a "cloud" of thermal neutron is formed around the detector. Studying of this thermal neutrons cloud was conducted using a standard gas-filled cylinder thermal neutron detector containing $BF_3$ gas. The detector has a diameter of 2 cm, the effective working length is about 17 cm, gas pressure is 800mm Hg, and the effectiveness of the thermal neutrons registration is about 13.5%.

The resulting spectrum of thermal neutrons is shown in Figure 4. The spectrum shows two peaks, which correspond to two types of nuclear reaction of neutron capture by the $^{10}B$ atom: the first type occurs when the total energy of the emitted alpha particle and the $^{7}Li$ ion is equal to 2.31 MeV (1.47 + 0.84), which happens in 94% of cases, and the second type of reaction occurs when the total energy of the emitted alpha particle and the $^{7}Li$ ion is equal to 2.79 MeV (1.78 + 1.01), which happens in 6% of cases.
In the case where the neutron capture takes place in a thin layer of the gas near the walls of the tube, either the alpha particle or the $^{7}Li$ ion is stuck in the tube wall, and the gas is ionized by either the alpha particle only, or by the $^{7}Li$ ion only.

Therefore, the spectrum shows two plots, respectively for the alpha particle with the energy of 1.47 MeV (94%) and for the $^{7}Li$ ion with the energy of 0.84 MeV (94%), for cases when the neutrons are captured near the tube wall.

From measurements with the $BF_3$ detector we have found that in the place where we put our multi-layered detector, the $^{241}Am$+Be source and two blocks of paraffin with a thickness of about 10 cm create a thermal neutron flux through the detector of the order of 150 n/cm$^2$ sec.

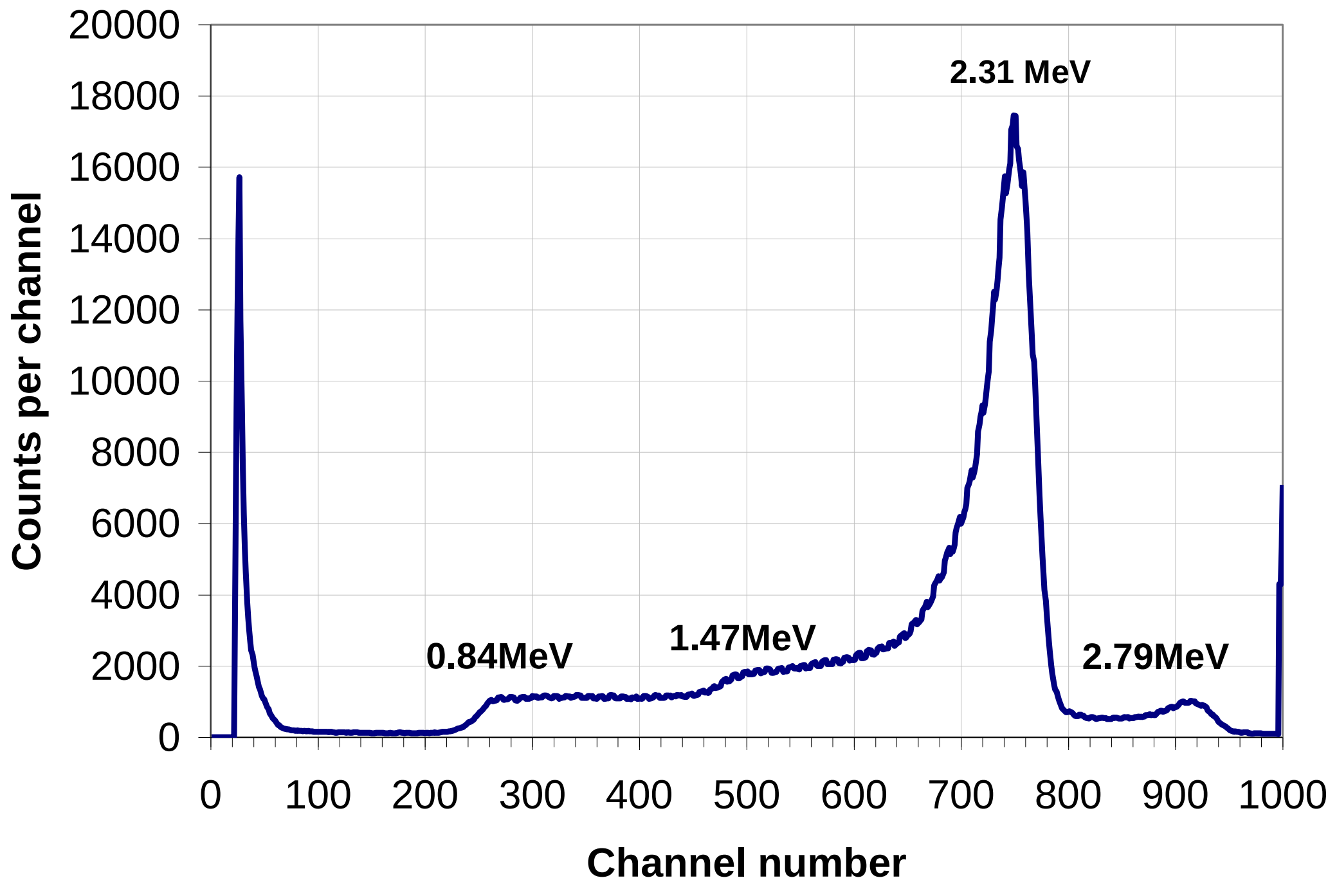

*Figure 4 Thermal neutrons spectrum detected by the tube detector $BF_3$ (2 cm diameter, 20 cm long) in the centre of measuring box, emitted by AmBe source + 2 paraffin blocks.*

## 4. Result of measurements

The spectra of signals and the intensity of the neutron counting were measured for detectors with different number of converter layers and with various effective area of these layers.
The measurements were performed in two stages:

The number of pulses coming from the detector was counted at the first stage with the Am+Be neutron source installed in the located on the top paraffin block. At the second stage the counting was performed without the source. In the latter case the level of the electronic noise and the background radiation was measured. Both measurements took the same time interval. The difference between these two measurements gave us the number of neutrons registered by our detector during a single measurement.

Based on the aforesaid, we obtained the following values of the efficiency of thermal neutron registration:

*Table 1 Technical parameters of fabricated multilayer thermal neutron detectors .*

| Number of layers (converter) in detector | Converter material | Single layer area [cm$^2$] | Thickness of a single layer [mg/cm$^2$] | Detection efficiency [%] |
|---|---|---|---|---|
| 2 | Boron carbide | 24 | 1.1 | 2.9 |
| 4 | Boron natural | 16 | 0.5 | 5.6 |
| 4 | Boron natural | 100 | 0.5 | 5.5 |
| 8 | Boron natural | 50 | 1.1 | 8 |
| 12 | Boron carbide | 16 | 0.7 | 10 |
| 16 | Boron natural | 40 | 0.5 | 12.5 |

Using of $^{10}$B-enriched (97%) boron as a neutron absorber must result in the growth of neutron registration efficiency by 5 times.

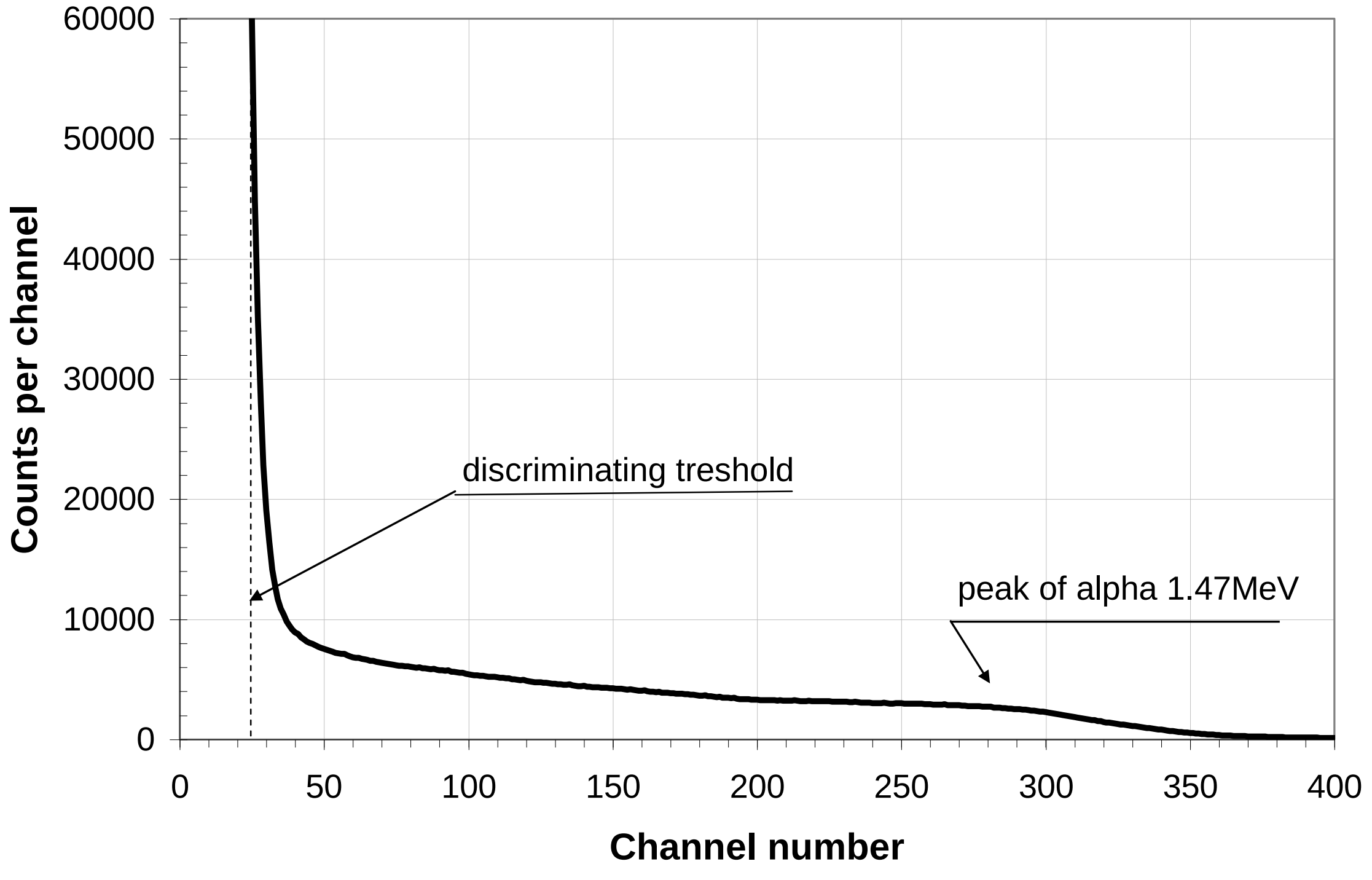

*Figure 5 Pulse height response of thermal neutrons obtained with multilayer (basic element 4-layers) detector*

Figure 5 shows the thermal neutron response of the multilayer (basic element 4-layers) composite detector (see Figure 2) measured in the air.

Additional characteristics of the measured sample including the obtained results can be found in Table 2.

*Table 2. Specifications of the multilayer (basic element 4-layers) detector and obtained thermal neutron detection results (measured in the air)*

| | | | | | |
|---|---|---|---|---|---|
| **Converter material** | **Nat. Boron** | **Amplification coefficient** | **$5 \cdot 10^5$** | **Count rate with source ($sec.^{-1}$)** | **680** |
| **Upper and bottom contacts** | **Fe** | **Thickness of the layer of air between the two converters** | **5mm** | **Count rate without source ($sec.^{-1}$)** | **6** |
| **Central contact** | **Cu** | **Voltage scale (mV/channel)** | **10** | **Calculated neutrons rate ($s^{-1}$)** | **674** |
| **One layer converter thickness ($mg/cm^2$)** | **1.1** | **Bias voltage (V)** | **600** | **Threshold channel** | **25** |
| **Area ($cm^2$)** | **92** | **Measurement time (s)** | **3000** | **Signal to noise ratio** | **113** |

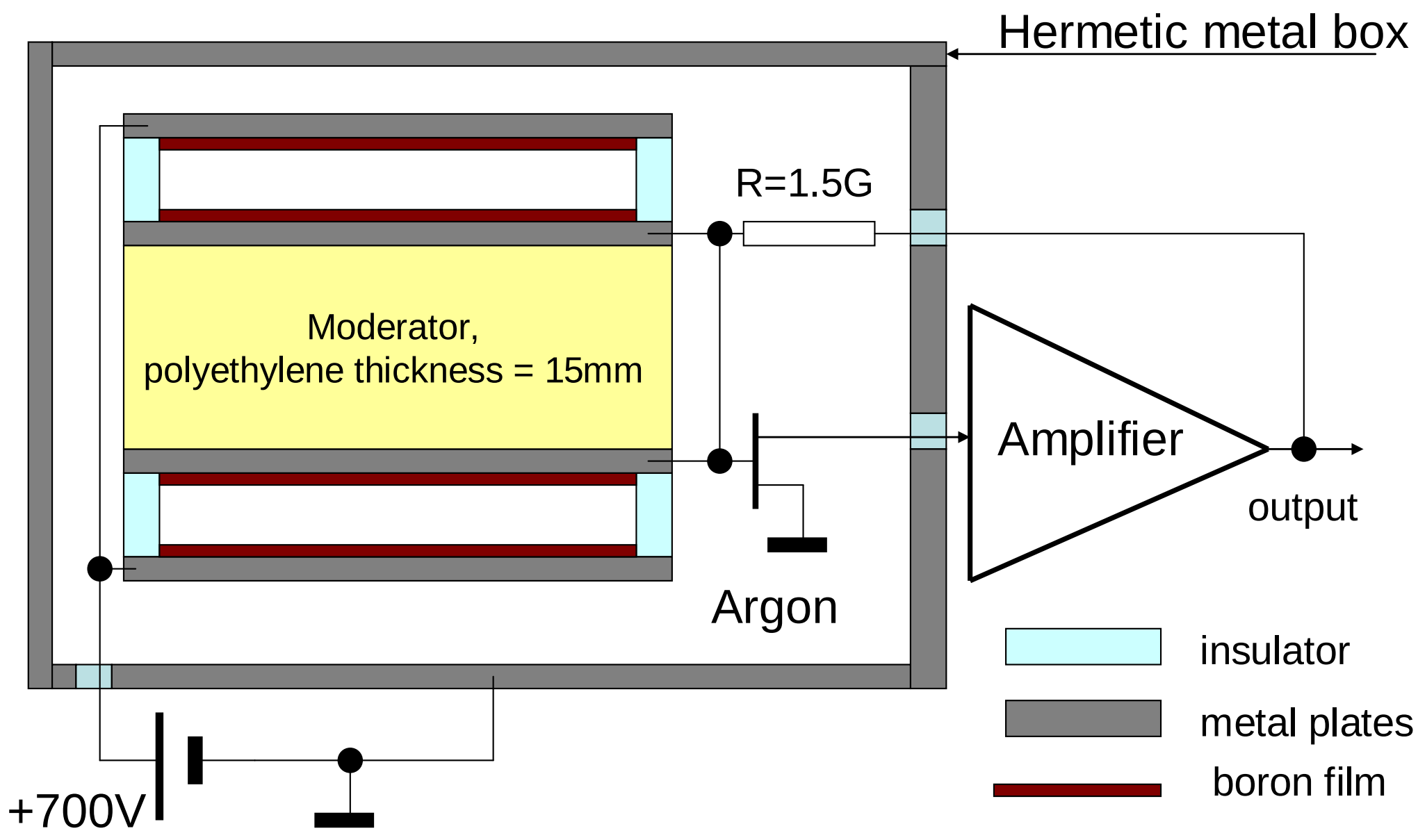

*Figure 6. 4-layers thermal neutron multilayer detector and electrical connection schema*

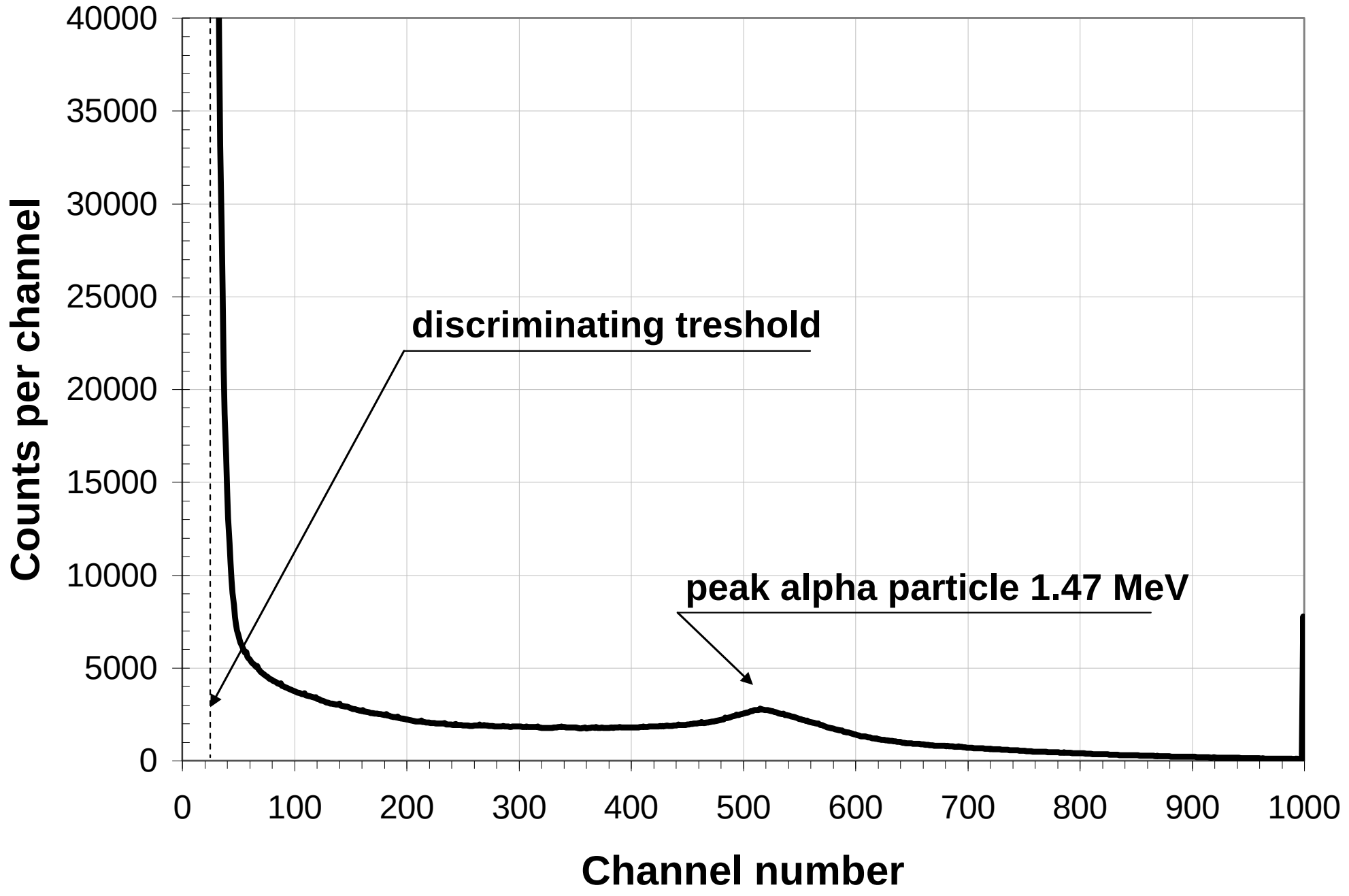

*Figure 7. Pulse height response of thermal neutrons obtained with 4-layer detector + polyethylene moderator.*

Figure 7 shows the thermal neutron response of this detector measured in argon. Additional characteristics of the measured sample including the obtained results can be found in Table 3.

*Table 3. Specifications of the multilayer (4-layers) detector and obtained thermal neutron detection results (measured in argon).*

| | | | | | |
|---|---|---|---|---|---|
| **Converter material** | **Nat. Boron** | **Thickness of the layer of argon between the two converters** | **6 mm** | **Count rate with source (sec.$^{-1}$)** | **840** |
| **Upper and bottom contacts** | **Fe** | **Amplification coefficient** | **$4 \cdot 10^5$** | **Count rate without source (sec.$^{-1}$)** | **14** |
| **Central contact** | **Fe** | **Voltage scale (mV/channel)** | **10** | **Calculated neutrons rate (s$^{-1}$)** | **826** |
| **One layer converter thickness (mg/cm$^2$)** | **0.5** | **Bias voltage (V)** | **700** | **Threshold channel** | **30** |
| **Area (cm$^2$)** | **100** | **Measurement time (s)** | **3000** | **Signal to Noise ratio** | **60** |

Additional testing was fulfilled to check the multilayer detector gamma-sensitivity. For this purpose there was fabricated 4-layer detector (see Figure 6) with geometry: electrode area 100 $cm^2$, natural boron layer thickness 0.5 $mg/cm^2$ and electrodes spacing 6 mm.

The $^{241}Am$ (activity 10 µCi) gamma source was mounted at the height of 2 cm above the upper electrode. This gamma source created 59.6 KeV photons flux ~$1.53 \times 10^6$/s through the lead aperture 5 mm diameter. The measurements were hold under the next conditions: amplification coefficient 400000, pulse shaping time 10 µs, voltage scale 10 mV/channel, bias voltage = 700V, measurement time 3000sec, and threshold channel 30. The received detection efficiency less than $5 \times 10^{-6}$ confirm our previous conclusion [2] of low gamma-sensitivity for multilayer thermal neutron detector detectors.

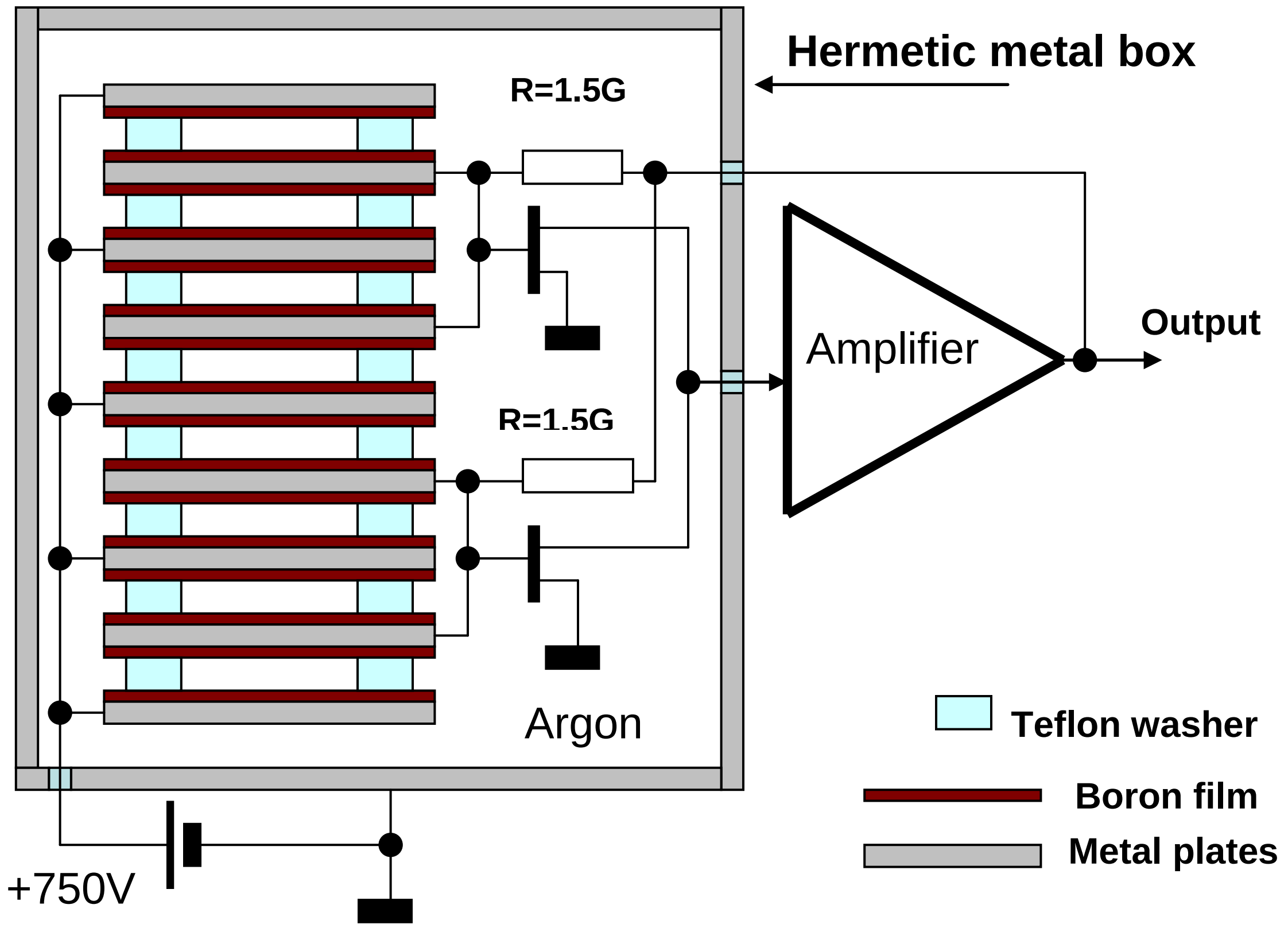

*Figure 8 16-layer thermal neutron detector and electrical connection schema.*

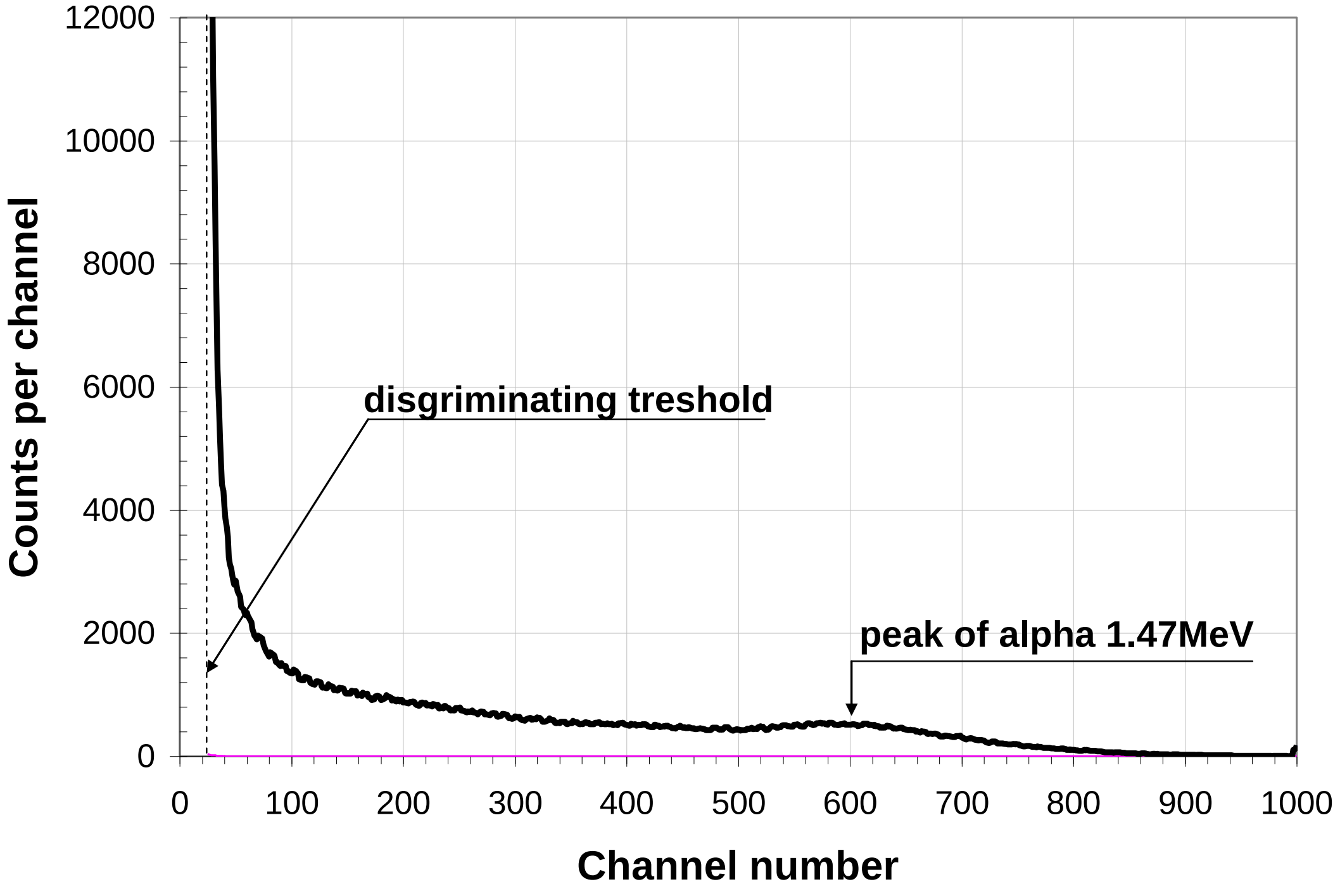

*Figure 9. Pulse height response of thermal neutrons obtained with multilayer (16-layers) detector*

Figure 9 shows the thermal neutron response of this detector measured in argon. Additional characteristics of the measured sample including the obtained results can be found in Table 4

*Table 4. Specifications of the multilayer (16 -layers) detector and obtained thermal neutron detection results (measured in argon).*

| | | | | | |
|---|---|---|---|---|---|
| **Converter material** | **Nat. Boron** | **Thickness of the layer of argon between the two converters** | **4 mm** | **Count rate with source (sec.$^{-1}$)** | **904** |
| **Upper and bottom contacts** | **Fe** | **Amplification coefficient** | **$5 \cdot 10^5$** | **Count rate without source (sec.$^{-1}$)** | **3** |
| **Central contact** | **Fe** | **Voltage scale (mV/channel)** | **10** | **Calculated neutrons rate (s$^{-1}$)** | **901** |
| **One layer converter thickness (mg/cm$^2$)** | **0.5** | **Bias voltage (V)** | **750** | **Threshold channel** | **25** |
| **Area (cm$^2$)** | **40** | **Measurement time (s)** | **1000** | **Signal to noise ratio** | **301** |

All three spectra (Figures 5, 7, 9) have clearly defined peak corresponding to the cases where alpha particles with the energy of 1.47 MeV are emitted from the converter surface and all of their energy is spent on the ionization of the gas-filled gap.

In cases where either alpha particles or $^{7}$Li ions are emitted from the depth of the convertor layer, they lose part of their energy in the converter and therefore we observe a continuous spectrum on the left side from the alpha peak. In cases where $^{7}$Li ions with energy of 0.84 MeV are emitted from the convertor surface there is not apparent peak in the spectrum because they are shaded by alpha particles signals which amplitudes are in the range of 0 -1.47 MeV.

The amplitude of electrical signals received from the detectors depends on the degree of charge collection in the ionization of gas-filled gaps. The amplitude increases with increasing electrical field applied to the gas-filled gaps. Also, the amplitude increases with the use of gases with a smaller amount of energy needed to create one electron – ion pair. Therefore, the amplitude of the signals in the argon is much larger than for the signals in the air. This is explained by the fact that oxygen presented in the air greatly affects the collection of electrons. However, the use of the air as a working gas greatly simplifies and reduces the cost of the detector and in some cases it is justified. Time of collection of electric charge in the gas-filled gap, and hence the duration of the pulse current recorded by electric circuit, decreases with increasing of the electrical field. For example, the time of collecting electrons in the air layer 4 mm thick and the electric field stronger than 200 V/mm decreases to 50 μs, while for argon that time decreases to 1 μs.

## 5. Summary and conclusion

Multilayer thermal neutron detector uses thin layers of absorbers containing the $^{10}$B isotope. If the detector is placed between the layers of the moderator, such as polyethylene, it can be used as a detector of fast neutrons with higher energies. Replacing the air by other gases such as argon, leads to a significant increase in the amplitudes of the signals from the detector, which increases the signal to noise ratio and allows to record signals corresponding to the lower energy alpha particles or $^{7}$Li ions. When charged heavy particles emitted from the depth of the neutron absorber layer, they lose part of their energy within this layer and only the remaining part of the energy is used for ionization of the gas layer.

The detector efficiency increases with the number of convertor layers in the detector.
The effectiveness of the detector with a single layer of the converter $^{10}$B isotope is about 6% for thermal neutrons [3,4,5]. Using of $^{10}$B isotope as a converter will allow our multi-layered detector to reach the efficiency obtained from the detectors using $^{3}$He gas, especially taking into account the actuality of the problem of replacing $^{3}$He detectors to other neutron detectors because of the existing shortage of $^{3}$He gas.

The multilayer thermal neutron detectors developed by us does not yield to $^{3}$He gas detectors in such basic parameters as signal count rate, signal amplitude, stability over time, signal to noise ratio, low sensitivity to gamma radiation. On the other hand, it is 5-10 times less expensive. The flat form of our detector and unlimited size of the working area are another advantages compared to $^{3}$He gas containing detectors. However, the main advantage of our detector consisting of individual pixels is the ability to acquire in real-time an image of the object through which the thermal neutrons emitted from the neutron generators or from nuclear reactors.

## References

[1] F. Knoll, Radiation Detection and Measurements, 2nd ed., Wiley, NY, p.483 (1989).
[2] M. Schieber, E. Mojaev, M. Roth, A. Zuck, O. Khakhan, A. Fleider ,Nucl.Instr. and Meth.**A607,** 634 (2009).
[3] R. Arnaldi1, E. Chiavassa1, A. Colla, P. Cortese, G. Dellacasa, N.De Marco, A. Ferretti, M. Gagliardi1, M. Gallio, R. Gemme, A. Musso, C. Oppedisano, A. Piccotti, F. Poggio, E. Scomparin and E. Vercellin J.Phys.: Conf. Ser. **41,** 384 (2006).
[4] M. Nakhostin*, M. Baba, T. Itoga, T. Oishi, Y. Unno, S. Kamada and T. Okuji **,** Protection Dosimetry. Advance Access originally published online on November 11, 2007, Radiation Protection Dosimetry **129,** 426 (2008).
[5] D.S..S.McGregor,S.M.Vernon,H.K.Gersch, S.M.Markham,S.J.Wojtczuk and D.K.Wehe IEEE Trans.Nucl.Sci.**47,** 1364 (2000).

* Corresponding author: Oleg Khakhan, e-mail - kholeg@cc.huji.ac.il